\def\C        {{$^{13}$C \/}}
\def\N        {{$^{14}$N \/}}
\def\NN       {{$^{15}$N \/}}
\def\NV        {{$^{14}$NV \/}}
\def\NNV       {{$^{15}$NV \/}}

\newcommand{\ee}[1]{\cdot10^{#1}}
\newcommand{\mr}[1]{\mathrm{#1}}
\newcommand{\unit}[1]{\,\mathrm{#1}}

\newcommand{\us}{\,\mu{\rm s}}

\newcommand{\muB}{\mu_{\rm B}}
\newcommand{\rtHz}{\sqrt{\mr{Hz}}}

\newcommand{\tc}{\tau_c}
\newcommand{\Dw}{\Delta\omega}

\newcommand{\MMd}{\langle\Delta\omega^2\rangle_\mr{d}}

\newcommand{\captionstyle}{\normalfont} 

\documentclass[prl,aps,twocolumn]{revtex4}

\usepackage{graphicx}
\usepackage{bm}
\usepackage{amsmath}
\usepackage{amssymb}
\usepackage{verbatim}
\usepackage[colorlinks=true, pdfstartview=FitV, linkcolor=blue, citecolor=blue, urlcolor=blue]{hyperref} 

\begin{document}

\global\emergencystretch = .1\hsize 

\title{Spin Properties of Very Shallow Nitrogen Vacancy Defects in Diamond}

\author{B. K. Ofori-Okai$^{1}$, S. Pezzagna$^{3}$, K. Chang$^{2}$, M. Loretz$^{2}$, R. Schirhagl$^{2}$, Y. Tao$^{1,2}$, B. A. Moores$^{2}$, K. Groot-Berning$^{3}$, J. Meijer$^{3}$, C. L. Degen$^{2}$}
  \email{degenc@ethz.ch} 
  \affiliation{
   $^1$Department of Chemistry, Massachusetts Institute of Technology, 77 Massachusetts Avenue, Cambridge MA 02139, USA.
   $^2$Department of Physics, ETH Zurich, Schafmattstrasse 16, 8093 Zurich, Switzerland.
   $^3$RUBION, Ruhr-Universitaet Bochum, Universitaetsstrasse 150, 44780 Bochum, Germany. }

\begin{abstract}
We investigate spin and optical properties of individual nitrogen-vacancy centers located within 1-10 nm from the diamond surface.  We observe stable defects with a characteristic optically detected magnetic resonance spectrum down to lowest depth.  We also find a small, but systematic spectral broadening for defects shallower than about 2 nm.  This broadening is consistent with the presence of a surface paramagnetic impurity layer [Tisler et al., ACS Nano {\bf 3}, 1959 (2009)] largely decoupled by motional averaging.  The observation of stable and well-behaved defects very close to the surface is critical for single-spin sensors and devices requiring nanometer proximity to the target.
\end{abstract}

\pacs{76.30.Mi, 75.70.Cn, 68.35.Dv}

\maketitle

Isolated defect spins in solids, such as phosphorus donors in silicon \cite{feher59} or the nitrogen-vacancy (NV) center in diamond \cite{jelezko06}, are considered important building blocks for future nanoscale devices, governed by quantum mechanics.  In pure materials, defects can be so well-decoupled from their solid host that spin states approach a stability normally only found in dilute atomic vapors, including coherence times of milliseconds to seconds \cite{balasubramanian09,tyryshkin11}.  Not surprisingly, atomic defects have over the last decade attracted increasing attention motivated by their potential for spin qubits in quantum information \cite{kane98,awschalom07} or for ultrasensitive magnetic detectors with nanometer spatial resolution \cite{degen08,maze08,balasubramanian08}.

The central challenge with many of these endeavors is to position the defect of interest in close proximity to other circuit elements while retaining their well-defined properties known from the bulk.  On the one hand, close proximity is required for strong enough coupling.  For example, for the direct coupling to nearby spin magnetic dipoles --- which scales as $r^{-3}$, where $r$ is distance --- efficient coupling is only achieved at nanometer separations.  Furthermore, for scanning magnetometry applications $r$ directly sets the attainable spatial resolution \cite{degen08}.  On the other hand, the coupling will almost always happen across a material interface and defects will have to be located within nanometers from a surface, potentially destabilizing the spin and limiting its usefulness.

Several mechanisms have been found or proposed to affect the stability of shallow defects.  For single donor spins in silicon, for example, the nearby Si/SiO$_2$ interface was shown to decrease spin coherence times even for donors tens of nanometer away due to paramagnetic impurities present at the interface \cite{desousa07}.  Other possible mechanisms include electric surface charge or strain fluctuations that may disturb defects through Stark and spin-orbit effects, or direct ionization \cite{hu06,rondin10}.  For nitrogen-vacancy centers in diamond, on the other hand, rather little is currently known about the spin's performance near the surface.  While functional defects have recently been reported in $<10\unit{nm}$ diameter nanocrystals \cite{tisler09,bradac10} and within $3-4\unit{nm}$ from bulk diamond surfaces \cite{grotz11} and coherence times $T_2$ of tens of $\us$ have been observed for defects at $\sim10\unit{nm}$ proximity \cite{gaebel06,maurer10,grinolds11}, neither a ``shallowest depth'' nor the involved destabilizing mechanisms are known.  Given the fundamental importance of surface proximity for applications, it appears imperative to experimentally explore the limits to stability of defects at very shallow depths.

Here we report a systematic study of the spin resonance properties of single NV defect centers down to a proximity of about $1\unit{nm}$.  Defects were produced by low energy ion implantation (0.4-5 keV) and investigated by optically-detected magnetic resonance (ODMR) spectroscopy.  We find well-behaved defects exhibiting a narrow electron spin resonance (ESR) spectrum and coherence times exceeding 10 microseconds down to the shallowest investigated depths.  We also observe extra line broadening for defects shallower than 2 nm.  This broadening is compatible with the presence of surface magnetic impurities that are mostly decoupled from the NV spin by motional averaging.  

\begin{figure}[t]
\centering
\includegraphics[width=0.50\textwidth]{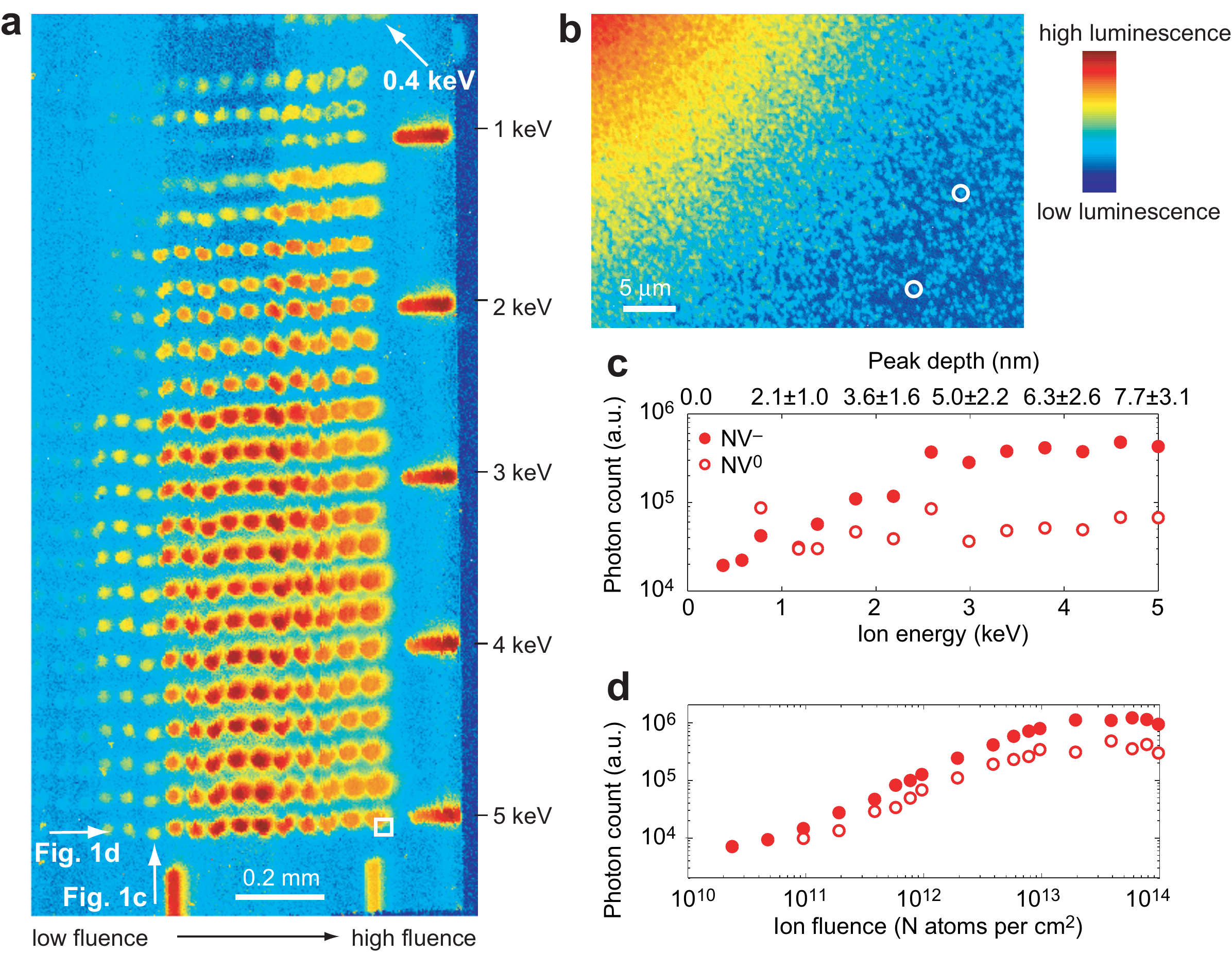}
\caption{\captionstyle (a) Photoluminescence intensity map of the diamond surface. Shallowest 0.4-keV defects are visible at the very top. Excitation wavelength was 532 nm and detection bandwidth 630-800 nm for all experiments except Fig. 1(c,d). (b) Close up of an implantation spot at $5.0\unit{keV}$ (white square in (a)). Some single NV centers are encircled. (c) Luminescence intensity due to NV$^-$ (full circles) and NV$^0$ (empty circles) as a function of ion energy at a fluence of $8\ee{11}\unit{cm^{-2}}$, measured using two pairs of filters and corrected for spectral overlap \cite{supplementary,rondin10}. (d) Luminescence intensity as a function of ion fluence at an energy of $5\unit{keV}$.}
\label{fig:fluorescence}
\end{figure}

A (100)-oriented single crystal of ultrapure diamond ($<5$ ppb N concentration, Element Six) was used as the sample for all experiments.  One sample face was implanted with $^{15}$N$^+$ or $^{15}$N$_2^+$ ions at a series of very low energies ($0.4-5\unit{keV}$, in steps of 0.2 keV) and fluences ($10^{10}-10^{14}\unit{N/cm^2}$) \cite{pezzagna10,implantation} (see Figure \ref{fig:fluorescence}).  To form NV centers, the sample was annealed for $2\unit{h}$ at $800\unit{^\circ C}$ and $p<2\ee{-7}\unit{mbar}$.  It is expected that nitrogen atoms do not diffuse at these temperatures because the activation energy is too high \cite{koga03,toyli10,supplementary}.  The sample was cleaned by boiling it for 24h under reflux in a 1:1:1 mixture of sulfuric, nitric and perchloric acid and thoroughly rinsed with purified water \cite{tisler09}.  This procedure is known to remove any residues (such as graphite) but to leave the diamond $sp^3$ bonding network intact.  Acid treatment also leaves a well-defined, oxygen-terminated reference surface \cite{tisler09}.  Additional details on implantation and sample preparation are given as Supplementary Material \cite{supplementary}.

We have performed a detailed inspection of the prepared diamond surface to validate the sample for later spin resonance measurements.  Surface roughness was determined by atomic force microscopy and was found to be very low ($x_\mr{rms} = 0.38\unit{nm}$, over a $300\times300\unit{nm^2}$ window) compared to the defect depth ($>1\unit{nm}$).  Angle-resolved x-ray photo-electron spectroscopy (ARXPS) was used to confirm oxygen termination of the surface and the absence of significant graphite residue.  The absence of $sp^2$ carbon was further corroborated by confocal Raman spectroscopy.  No difference was found between implanted and non-implanted regions.

A photoluminescence intensity map of the sample is shown in Figure \ref{fig:fluorescence}(a).  Most prominently, we observe that optically bright NV centers are visible down to the lowest implantation energy (0.4 keV).  We have estimated the depth of these defects using stopping range of ions in matter (SRIM) Monte Carlo simulations \cite{ziegler08,supplementary}; for example, an energy of 0.4 keV corresponds to a peak depth of $1.1\pm0.6\unit{nm}$ (see scale in Fig.\ref{fig:fluorescence}(c)).  The advantage of SRIM calculations, which is the primary method to determine defect depths in this study, is that they give suitable results of ion implantation depths over a wide energy range, including very low implantation energies \cite{honicke10}.  This is important given the current lack of a precise experimental method to directly measure surface proximity.  The drawbacks of SRIM calculations are that the results are inherently statistical (which is addressed below by collecting statistics on many defects), and important biasing effects like channeling are not considered.  We have analyzed these effects for our study (see Supplementary Material).  In particular we find that ion channeling, which could lead to depth underestimation by about $2\times$ \cite{toyli10,nordlund}, does not occur for \N energies below 0.6 keV and is only important towards higher energies \cite{supplementary}.  The lowest energies, where channeling is absent, are the most relevant in this study.

Figures \ref{fig:fluorescence}(c,d) provide additional photoluminescence data that further corroborates this picture.  We have measured the total photoluminescence intensity both as a function of energy and ion fluence and determined the relative concentrations of NV$^-$ and NV$^0$ centers.  In good agreement with earlier studies carried out at higher energies \cite{pezzagna10} we find a monotonic decrease in total photoluminescence with decreasing energy.  This decrease has been attributed to the vacancy-limited formation of NV centers \cite{pezzagna10}.  Since the decrease appears to be mostly due to a reduction of NV$^-$ one could conjecture the presence of a depth threshold below which the negative charge state becomes unstable \cite{rondin10}.  We have not, however, observed any photobleaching and we have only seen a few rare cases of fluorescence intermittency among the investigated single centers \cite{bradac10} that would support such a threshold.  The presence of a threshold is also incompatible with the rapid changes in the spin resonance linewidth that we see for the lowest energies (see below).  

We now turn to the core part of this Letter which is a study and analysis of electron spin resonance (ESR) spectra as a function of defect depth.  ESR measurements are carried out using optically detected magnetic resonance spectroscopy \cite{gruber97}.  For these measurements, the fluorescence intensity from single, isolated NV center is collected while slowly sweeping an auxiliary cw microwave field across the spin resonance ($\sim 2.8\unit{GHz}$) of the electronic ground state.  Resonant microwaves induce transitions between the $m_s=0$ and $m_s=+1$ (or $m_s=-1$) spin sublevels and lead to an up to $30\%$ reduction in fluorescence.  We use this feature to map out ESR spectra of single defects and to measure their linewidth and coherence properties.

\begin{figure}[t]
\centering
\includegraphics[width=0.50\textwidth]{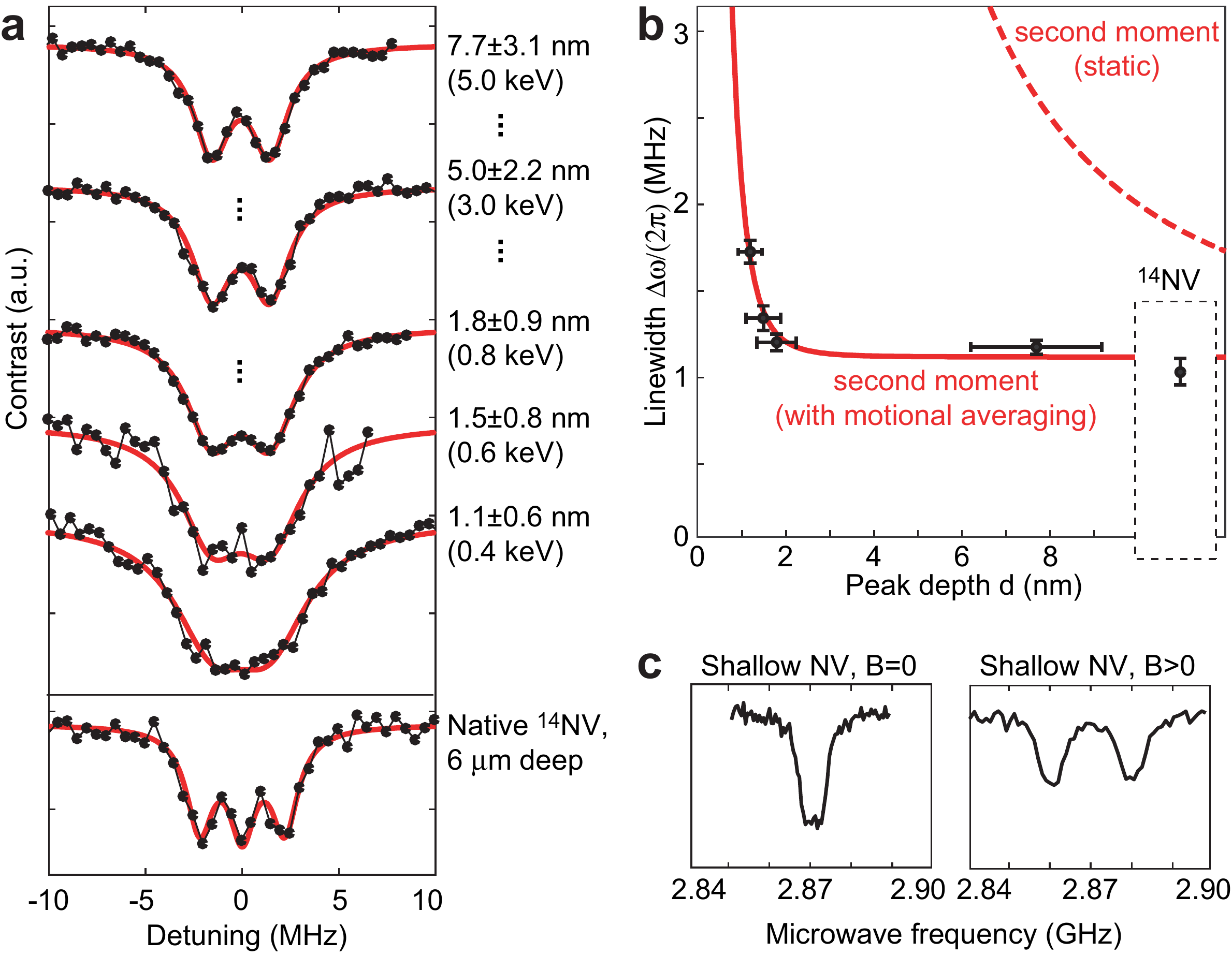}
\caption{\captionstyle Electron spin resonance measurements on shallow NV defect spins. (a) Representative spectra (out of $>30$ total) showing increased line broadening at shallow depths.  Black dots are experimental points and red solid lines are Lorentzian fits.  Implanted $^{15}$NV (nuclear spin $I=1/2$) are distinguished from native \NV ($I=1$, $99.6\%$ natural abundance) by the different hyperfine manifold.  The magnetic bias field is about 20 gauss. (b) ESR linewidth $\Dw/2\pi$ (half width at half height) plotted against surface proximity $d$.  Black dots are experimental values obtained from many separately fitted curves such as the ones shown in a). Error bars denote standard error.  The solid red line is a fit to the black dots based on Eq. (\ref{eq:mmdnarrow}) including motional averaging.  The dashed red line is the static second moment [Eq. (\ref{eq:mmd})] shown for comparison.
The linewidth of reference \NV is also shown.  (c) Spectra of a 1.1-nm defect at zero field (left) and at 3.5 gauss bias field (right) rule out the presence of significant surface strain or charge (see text).
}
\label{fig:odmr}
\end{figure}
Figure \ref{fig:odmr}(a) collects a series of representative ESR spectra taken on NV centers at different depths.  As a key feature we observe increased line broadening as NV spins are located closer to the surface:  For defects deeper than 1.8 nm (0.8 eV), there is a clear hyperfine splitting due to the \NN nuclear spin, but for the shallower defects at 1.5 nm (0.6 keV) and 1.1 nm (0.4 keV) the resonances become broad and the hyperfine doublet is barely visible or entirely unresolvable.  This picture of broadened lines was consistent among recorded spectra ($>30$ in total); for example, we did not find any 0.4 keV defects with a resolved hyperfine splitting, while most 0.8 keV and virtually all deeper defects showed a clear hyperfine doublet.

We have quantitatively analyzed the linewidth for the shallowest defects (where the pronounced changes are seen) by collecting and fitting a number of individual spectra and averaging the resultant linewidth parameter [Fig. \ref{fig:odmr}(b)].  While we have taken additional spectra at other (higher) energies that all show clear hyperfine doublets, these spectra do not have statistical significance and are not included in the figure.  Spectra are recorded at a single fluence ($8\ee{11}\unit{cm^{-2}}$) and on defects that lie at the perimeter of an implantation dot, where the density is low enough to optically isolate individual NV centers and residual dipolar broadening by N donor electronic spins can be excluded \cite{supplementary}.

Several control measurements were carried out to ensure that the observed line broadening is indeed a result of surface proximity.  A number of native \NV spectra was recorded at each investigated implantation spot by focusing slightly into the bulk in order to verify that broad lines were a property of the defect, and not, e.g., the sample or experimental parameters.  We have also measured a few spectra at higher fluence (up to $1\ee{13}\unit{cm^{-1}}$) and found that the line broadening did not change between implantation spots of the same energy but different fluence.  Moreover, no increase in \NV density is seen on or near implanted areas which eliminates the possibility that \NV centers were formed from vacancies created during \N implantation, in agreement with previous reports \cite{toyli10}.  Finally, we did not observe a line splitting at zero magnetic bias field [Fig. 2(c)], a signature for the presence of electric fields \cite{dolde11}, thereby excluding the presence of significant surface strain or charge.

\begin{figure}[t]
\centering
\includegraphics[width=0.30\textwidth]{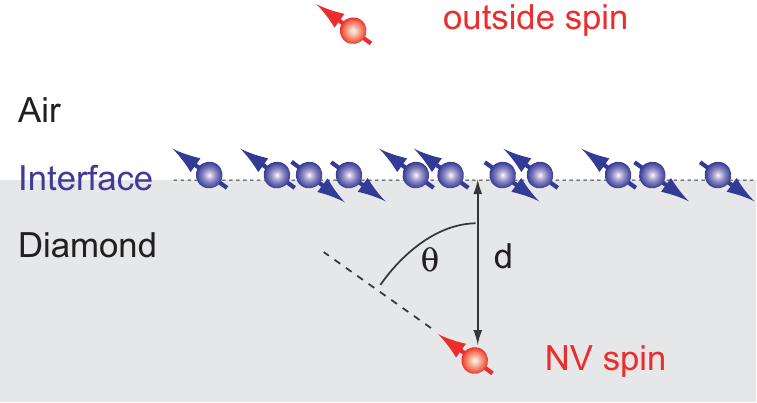}
\caption{\captionstyle  Sketch of a nitrogen-vacancy defect spin near a paramagnetic surface impurity layer, as described in the text.  $d$ and $\theta$ denote distance and orientation, respectively, of the defect spin to the interface plane, and quantization is along the NV axis.  A potential target spin, representative for sensing applications, is also shown.}
\label{fig:model}
\end{figure}

In the following we attempt to explain the surface-induced line broadening by the presence of paramagnetic impurities.  Surface impurities have been found at substantial density for clean, oxygen-terminated nanodiamonds \cite{tisler09}. These nanodiamonds were milled from larger crystals and underwent the same surface cleaning procedure, and are thus expected to exhibit a surface chemistry very similar to our diamond substrate.  We can model surface impurities by assuming a homogeneous, two-dimensional dipolar bath of electron spins ($S=1/2$) with an areal density of $\rho_A\approx 10\unit{spins/nm^2}$ \cite{tisler09}, similar to the sketch in Figure \ref{fig:model}.  In the following, we use the theory of moments \cite{slichter90}, but note that a parallel framework has been developed for $T_1$ and $T_2$ values in the context of paramagnetic impurities in the Si/SiO$_2$ interface \cite{desousa07}.  The second moment $\MMd$ ($\MMd \approx \Dw_\mr{d}^2$, where $\Dw_\mr{d}$ is linewidth) of the NV spin resonance at a distance $d$ from the spin bath is found to be 
\begin{eqnarray}
\MMd & = & \frac{3 \hbar^2 \mu_0^2 \rho_A \gamma^4}{2048 \pi d^4} \times (3+2\cos^2\theta+3\cos^4\theta),
\label{eq:mmd}
\end{eqnarray}
where $\mu_0=4\pi\ee{-7}\unit{Vs/(Am)}$, $\gamma/(2\pi) = 2.8\ee{10}\unit{Hz/T}$ is the surface electron gyromagnetic ratio, and $\theta$ is the angle between the NV axis and surface normal \cite{supplementary}.  For an NV center near a (100) surface ($\theta = 54.7^\circ$) the angular factor in (\ref{eq:mmd}) is $3+2\cos^2\theta+3\cos^4\theta = 4$.  The resonance linewidth that can be directly compared to the experiment is then $\Dw = \sqrt{ \Dw_0^2+\MMd}$, where $\Dw_0$ is the intrinsic linewidth (here $\Dw_0/(2\pi) = 1.1\pm0.1\unit{MHz}$).

Equation (\ref{eq:mmd}) describes the second moment of a quasi-static spin bath that does not fluctuate on the time scale of the NV dephasing time $T_2^\ast\approx\Dw^{-1/2}$.  This assumption is likely invalid for surface spins that are not protected by the diamond matrix.  For example, spin-lattice relaxation times $T_1$ observed for paramagnetic impurities in amorphous carbon \cite{barklie00} and sintered detonation nanodiamonds \cite{baranov11} are on the order of nanoseconds.  Following the work by Kubo and Tomita \cite{kubo54} we can calculate a modified second moment that takes into account fast fluctuations (''motional averaging''),
\begin{equation}
\MMd' = (\MMd \tc)^2, \quad\quad{(\tc \ll \MMd^{1/2})}
\label{eq:mmdnarrow}
\end{equation}
where $\tc\approx T_{1,\mr{surface\ spins}}$ is the correlation time of fluctuations. 

We can compare the model represented by Eqs. (\ref{eq:mmd}) and (\ref{eq:mmdnarrow}) through a fit to the experimental data in Figure \ref{fig:odmr}(b).  The three experimental parameters are the surface spin density $\rho_A$, the distance $d$, and the correlation time $\tc$.  Assuming $\rho_A=10\unit{spins/nm^2}$ \cite{tisler09} and $d$ from the SRIM calculation \cite{supplementary}, one obtains a correlation time $\tc<1\unit{ns}$.  This $\tc$ would be rather fast compared to the above literature values \cite{barklie00,baranov11} and typical organic radicals.  As neither $\rho_A$ nor $d$ are accurately known, our value for $\tc$ is at best approximate.  For example, if we allow $\rho_A$ to vary between $0.1-10\unit{spins/nm^2}$ and assume an underestimation of depth $d$ by up to $2\times$, then the range of compatible $\tc$ varies between $0.1-100\unit{ns}$.  Further evidence for a correlation time $\tc$ in the nanosecond range comes from relaxation time measurements on selected shallow defects where we observe $T_1\gg T_{1\rho}\approx T_2$ \cite{loretz}. 

\begin{figure}[t]
\centering
\includegraphics[width=0.45\textwidth]{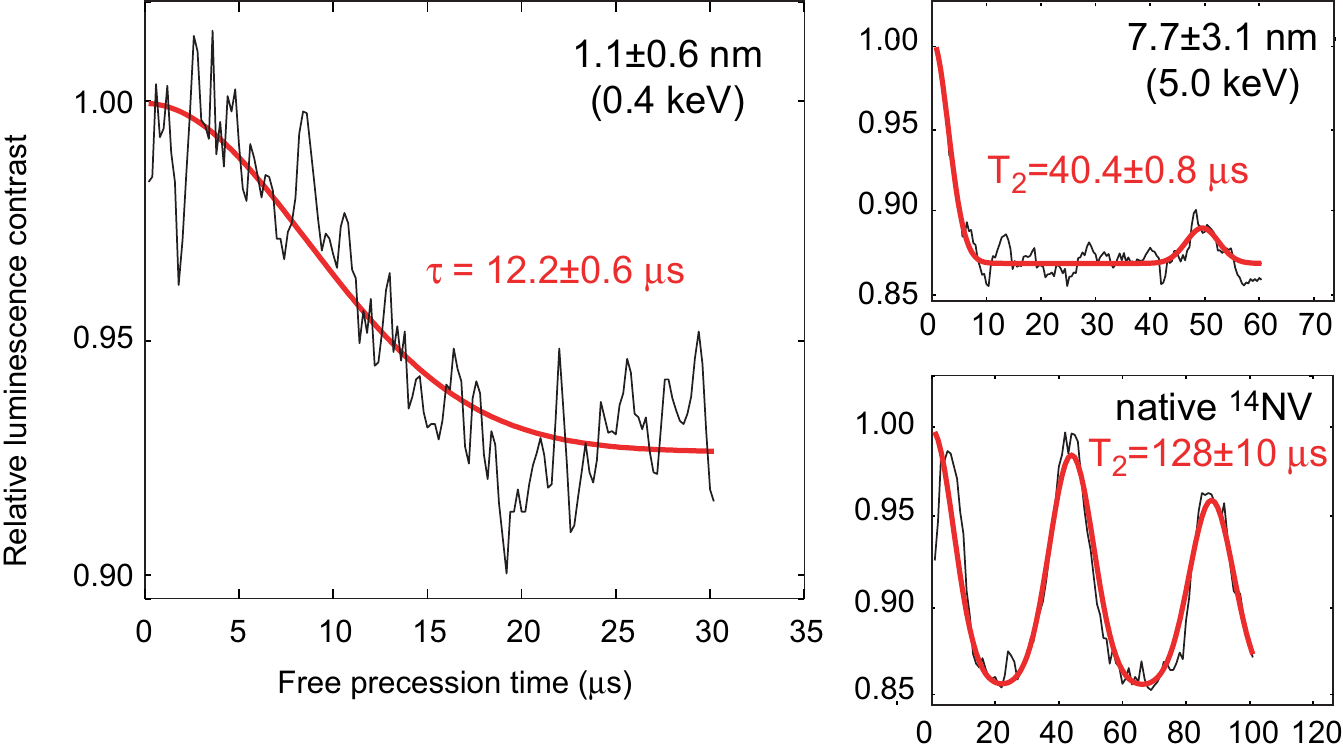}
\caption{\captionstyle Hahn-echo decay of a 1.1-nm \NNV center and, for comparison, for a 7.7-nm \NNV and a native \NV defect. Black curves are experimental data and red lines are fits.  The fit function is given in \cite{supplementary}.}
\label{fig:echo}
\end{figure}

Finally, we have also measured echo decay times of several 0.4 keV (1.1 nm) defects to establish a lower bound for the coherence times $T_2$ of very shallow spins.  A representative Hahn echo-decay curve with an echo decay time of $\tau = 12\unit{\us}$ is shown in Figure \ref{fig:echo}; other defects at the same depth showed $\tau$ values between 7 and $12\unit{\us}$.  We note that the echo decay profile is Gaussian, which is expected if decay is caused by the slowly fluctuating \C nuclear spin environment.  If echo decay were set by rapidly fluctuating surface impurities, then one would rather expect an exponential decay.  Consequently, one can conclude that surface spins are not the dominant source of decoherence and that the limit on $T_2$ implied by surface impurities is $\overset{>}{\sim}10\unit{\us}$.

Given the shallow depths of the investigated NV centers it is instructive to extrapolate their magnetic moment sensitivity, which is the key figure of merit for the sensing of external spins and future applications to nanoscale magnetic resonance imaging and spectroscopy \cite{degen08}.
For example, taking an echo decay time $\tau\sim12\unit{\us}$, a photon count rate of $C=0.0018\unit{photons/shot}$ and an optical contrast between $m_s=0$ and $m_s=\pm1$ states of $\epsilon=7\%$ (1.1-nm defect in Fig. \ref{fig:echo}), we find an optimal ac magnetic field sensitivity of $B_\mr{min} \approx (0.5\pi\gamma\epsilon\sqrt{\tau C})^{-1} \sim 2.2\unit{\mu T/\rtHz}$ \cite{taylor08}.  Here, the relevant (most susceptible) ac frequency is set by the inverse of the echo duration, i.e., tens of kHz.  For a magnetic moment located directly on the surface and taking into account the angle of the NV spin, this sensitivity equates to a minimum detectable magnetic moment of $\mu_\mr{min} \sim B_\mr{min}/[(0.96\times\mu_0\mu_B)/(4\pi d^3)] \sim 0.003\unit{\mu_B/\rtHz}$, where $\mu_B$ is the Bohr magneton.
%
For dc signal detection, the corresponding magnetic field and moment sensitivities are $B_\mr{min} \approx 8\Dw_0/(3\sqrt{3}\gamma\epsilon\sqrt{I_0}) \sim 19\unit{\mu T/\rtHz}$ and $\mu_\mr{min} \sim 0.03\unit{\mu_B/\rtHz}$, respectively, where $I_0=2\ee{3}\unit{photons/s}$ is the cw photon count rate and $\epsilon = 11\%$ [Fig. \ref{fig:odmr}(a)].
%
Even if our depth was underestimated by a factor of $2\times$ \cite{supplementary}, $\mu_\mr{min}$ would still equate to $0.03\unit{\mu_B/\rtHz}$ (ac) and $0.2\unit{\mu_B/\rtHz}$ (dc), respectively.

In conclusion, we have investigated spin and optical properties of single nitrogen vacancy defects in diamond at very shallow depths.  Functional defects are found down to about 1 nm, and significant broadening of the electron spin resonance is only observed for defects $<2\unit{nm}$.  This surface stability is unmatched by other solid-state spin systems, such as phosphorus donors in silicon or semiconductor quantum dots, and is a key requirement for a number of anticipated quantum and sensing applications.  In particular, we have inferred a sensitivity to outside magnetic moments (such as surface electron and nuclear spins) that extends down to $<0.01\unit{\muB/\rtHz}$.  To the best of our knowledge this is the best magnetic moment sensitivity demonstrated for a general-purpose magnetic sensor to date \cite{degen08nnano}.  If combined with the imaging capabilities of a scanning probe apparatus \cite{chernobrod05,degen08}, this sensitivity will enable the direct mapping of nuclear spins in molecules and thin films with chemical specificity and nanometer spatial resolution.

The authors gratefully acknowledge financial support through the NCCR QSIT, a competence center funded by the Swiss NSF, through SNF Grant No. $200021\_137520 / 1$, and through the Volkswagen Stiftung.  We thank G. Balasubramanian, H. Balch, J. Hodges, L. Robledo, and C. Ryan for support in constructing the optical setup, and A. Rossi and J. Stadler for help with the surface inspection.


\begin{thebibliography}{}
\bibitem{implantation}
	The use of molecular N$_2^+$ ions was necessary to access the lowest energies ($\leq 2.4\unit{keV}$).  No difference has been found between N$^+$ or N$_2^+$ implanted defects neither in this nor in an earlier study (Ref. \cite{pezzagna10}).  It is assumed that as the N$_2^+$ ion impinges the surface the molecule breaks, sending two N atoms into the bulk.
\end{thebibliography}
\end{document}




\begin{center}
{\Large Supplementary Material to the Manuscript\\ ``Spin Properties of Very Shallow Nitrogen Vacancy Center in Diamond''}
\vspace{0.5cm} \\
B. K. Ofori-Okai$^{1}$, S. Pezzagna$^{3}$, K. Chang$^{2}$, M. Loretz$^{2}$, R. Schirhagl$^{2}$, Y. Tao$^{1,2}$, B. A. Moores$^{2}$, K. Groot-Berning$^{3}$, J. Meijer$^{3}$, C. L. Degen$^{2}$ \\
\end{center}

{\small\noindent
$^1$Department of Chemistry, Massachusetts Institute of Technology, 77 Massachusetts Avenue, Cambridge MA 02139, USA. \\
$^2$Department of Physics, ETH Zurich, Schafmattstrasse 16, 8093 Zurich, Switzerland. \\
$^3$RUBION, Ruhr-Universitaet Bochum, Universitaetsstrasse 150, 44780 Bochum, Germany. \\
}


\section{Ion implantation and annealing}

\subsection{Ion implantation}
\begin{figure}[h]
\centering
\includegraphics[width=0.95\textwidth]{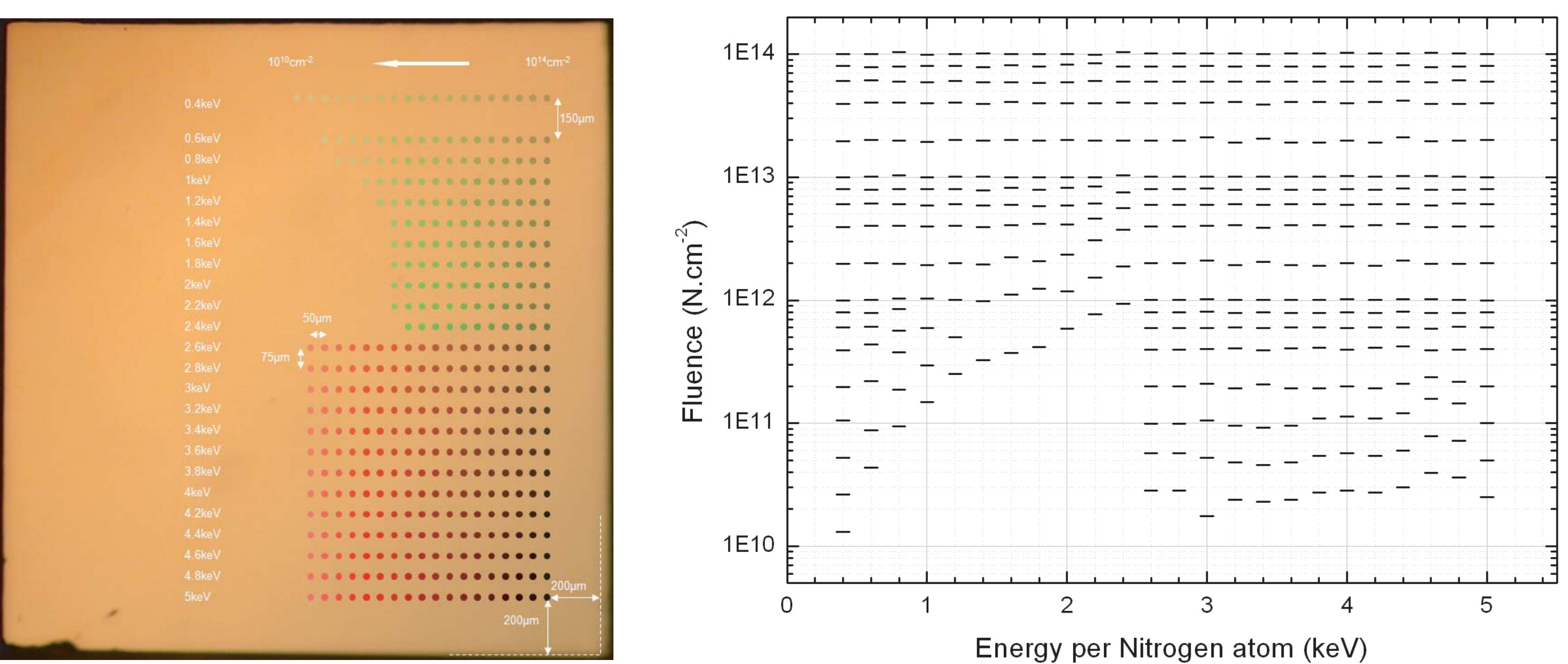}
\caption{\captionstyle (left) Spatial map of ion implantation. (right) Energies and fluences of the various spots.}
\label{fig:implantation}
\end{figure}

Ion implantation was carried out with a 5 keV gas-source ion gun (SPECS) combined with a Wien mass filter (E$\times$B).  Ion beam direction was normal to the diamond surface [a (100) surface] with a precision of better than $\pm 1^\circ$.  Spot size was controlled by a $25\unit{\um}$ aperture.  Fluence was measured via the ion current, and adjusted in a way that each spot received the same number of ions.  Thus, even if spots slightly vary in shape, the number of ions impacting the larger area of the spot is the same.  Miscalibration of the fluence is the largest error in the photoluminescence intensities presented in Fig. 1; in particular, it is most likely responsible for the low photoluminescence of the 1-keV row.  Figure \ref{fig:implantation} shows a map of the implantation pattern, energies, and fluences that were used on the investigated sample.  A detailed description of the ion implantation apparatus is given in Ref. \cite{meijer08}.

\subsection{Annealing}
Annealing was carried out in a AJA sputtering machine that allowed heating the diamond sample to $800^\circ$C while maintaining a pressure $<2\ee{-7}\unit{mbar}$ over the entire annealing duration of 2h, followed by a slow overnight cool-down.  It is assumed that this pressure is low enough to prevent oxidative etching of the top diamond layers.  It is also assumed that nitrogen atoms do not diffuse under these conditions.  Diffusion coefficients for N in diamond have been measured at higher temperatures in the context of geology \cite{koga03} and were found to follow an Arrhenius relationship,
\begin{equation}
D = 9.7\ee{-8}\unit{m^2/s} \ e^{\frac{-6.0\unit{eV}}{\kT}}.
\end{equation}
At $T=800\unit{^o C}$, the diffusion coefficient is $D=3\ee{-35}\unit{m^2/s}$.  The diffusion length for this $D$ and $t=10^4\unit{s}$ is $L=\sqrt{D t}=6\ee{-16}\unit{m}$.  This is much smaller than interatomic distances ($\sim10^{-10}\unit{m}$).  It is possible that the vacancies created around an implanted N atom will lower the activation energy for diffusion, however, such diffusion would be confined to the local area of the N defect and not significantly alter the N atoms's position.

\subsection{Exclusion of near-surface \NV formation}
There is a possibility that near-surface NV centers may be formed from native \N defects due to the vacancies created during ion implantation.  A signature of such (unwanted) \NV formation would be an enhanced density of \NV centers on or near implantation spots.  We have not observed such an increase with any of the implantation spots investigated by ODMR.  From this we conclude that no significant number of near-surface NV centers is formed from native \N defects.  This is plausible given the low intrinsic density of native \N of less than $<5$ppb \cite{edmonds12}.  It is also in agreement with earlier studies on N implantation into diamond \cite{toyli10}.
 

\section{Estimation of implantation depth for low energy ion implantation into diamond}

\begin{figure}[h]
\centering
\includegraphics[width=0.55\textwidth]{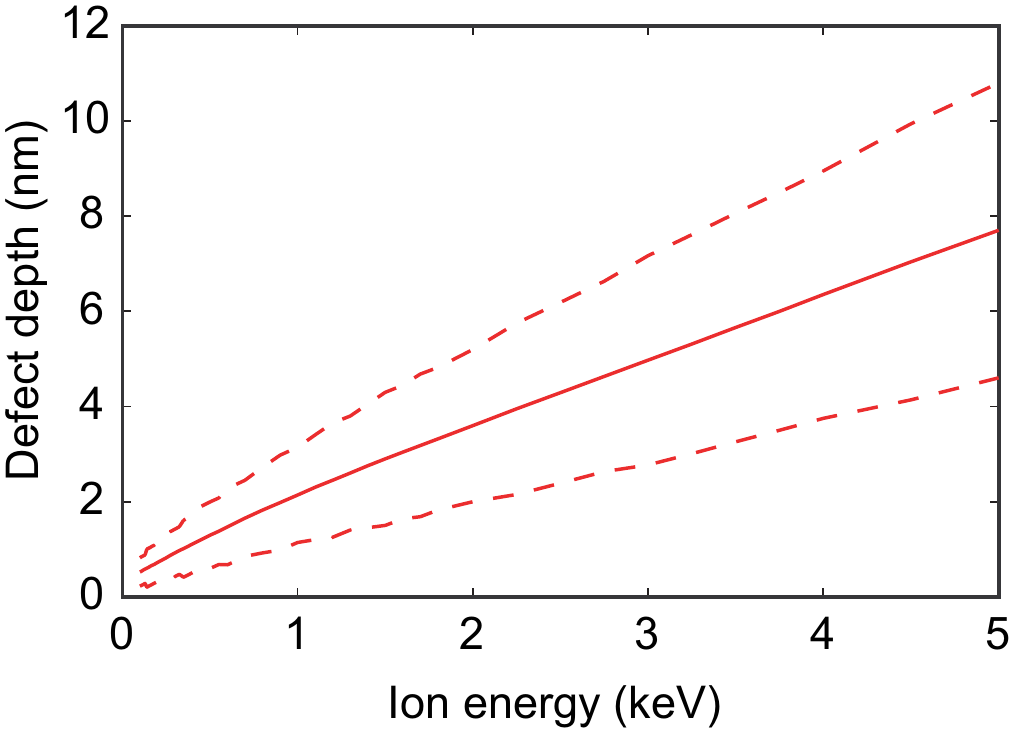}
\caption{\captionstyle SRIM Monte Carlo simulation for implantation of N ions into diamond. Solid line is peak depth (median) and dashed lines are $\pm1$ standard deviation.}
\label{fig:srim}
\end{figure}

The depth of defects created by ion implantation depends on the ion species, kinetic energy, and the target material.  A widely used approach to calculate defect depth is the stopping range of ions in matter (SRIM) Monte Carlo simulation \cite{ziegler08}.  It has been shown that SRIM simulations give suitable results of ion implantation depths over a wide energy range, including very low implantation energies \cite{honicke10}.  SRIM is the main approach used to estimate defect depths for our study.  Figure \ref{fig:srim} shows the calculated defect depth (peak depth) plotted against the ion energy for N ions in the range $0-5\unit{keV}$.

The SRIM simulation does however not consider two potentially important effects that influence implantation depth, namely, aperture scattering and ion channeling.  
These effects are discussed in the following.

\subsection{Aperture scattering}

Proximity effects due to scattered ions on the mask are a well known problem in semiconductor industry.  The scattered ions both reduce the lateral resolution and lead to a depth underestimation due to loss of kinetic energy by scattering.  The fraction of scattered ions depends on the aperture size, material, shape and the ion species and energy \cite{meijer08}.  An upper limit to the fraction of scattered ions is given by
\begin{equation}
f_\mr{scatt} = \left(1+\frac{\Delta R}{R_\mr{ap}}\right)^2-1 \approx \frac{2\Delta R}{R_\mr{ap}},
\end{equation}
where $R_\mr{ap}$ is the radius of the aperture and $\Delta R$ the enlargement due to ions that can penetrate the aperture and are scattered.  For our system $R_\mr{ap}=25\unit{\um}$ and $\Delta R\approx 5.8\unit{nm}$ (implantation depth of 5 keV N ions into a Pt aperture curved at $30^{o}$) the fraction is $f_\mr{scatt}\approx 10^{-3}$.  For the lowest energies used in our experiments the scattering fraction is reduced by an order of magnitude.  In addition, the aperture is further demagnified by an electrostatic lens to $20\unit{\um}$ and filtered with a second $200\unit{\um}$ aperture right in front of the lens.  This secondary aperature absorbs ions strongly scattered at the main $25\unit{\um}$ aperature.  Thus, we expect no significant impact of aperature scattering on ion implantation depth.

\subsection{Channeling}
Ion channeling describes a kinetic ion ``traveling'' along crystal planes, resulting in a reduced number of collisions. Channeling leads to an underestimation of defect depth \cite{derry82,smith91,toyli10}.  The relevant parameters concerning ion channeling are the angle between the ion beam and the crystal planes, and ion energy.  For a particular crystal orientation, only ions impinging within a certain acceptance angle can channel along crystal planes.  This angle is called the critical angle.  Furthermore, channeling does not occur below a certain critical kinetic energy threshold.

\begin{figure}[h]
\centering
\includegraphics[width=0.95\textwidth]{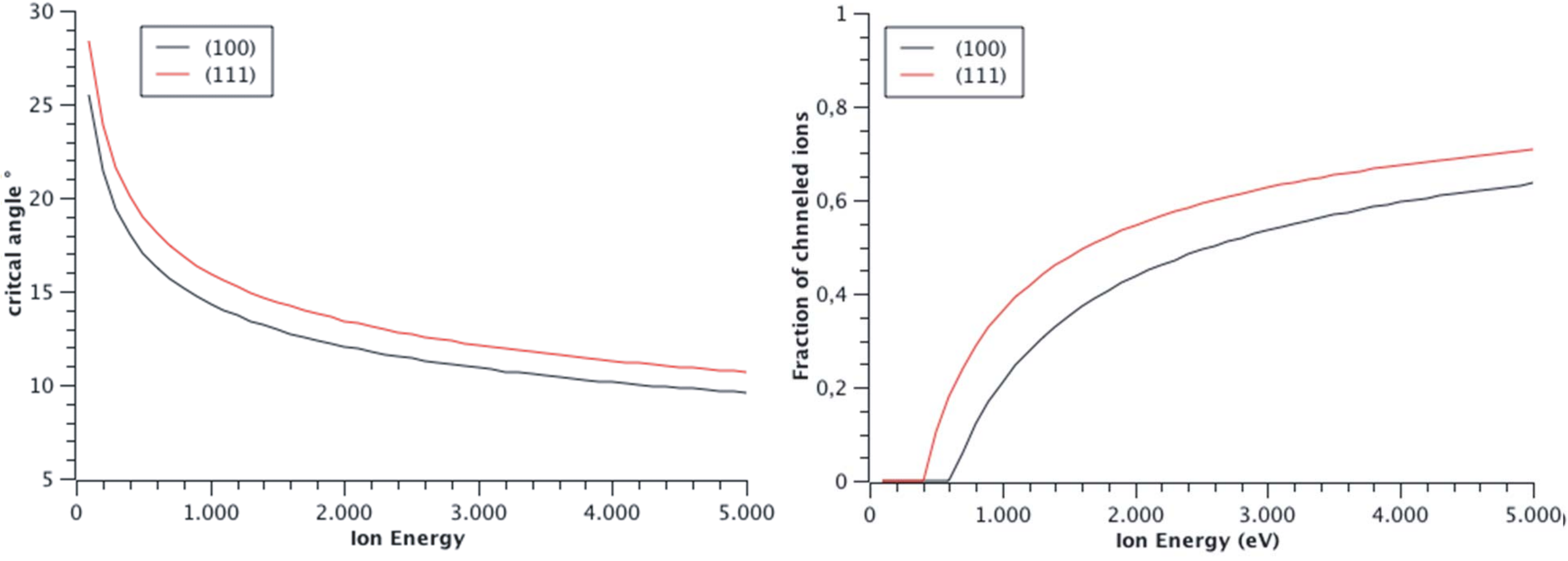}
\caption{\captionstyle (a) Critical angle and (b) Channeling fraction as a function of ion energy for N implantation into diamond. The critical energy is 0.6 keV for implantation into the (100) surface.}
\label{fig:channeling}
\end{figure}
We have calculated the critical angle and critical energy for N implantation into (100)-oriented diamond based on the model by Lindhart\cite{lindhard64,redinger11}. The results of these calculations are plotted in Figure \ref{fig:channeling}.  Fig. \ref{fig:channeling}(a) shows that the critical angle becomes large at low energies and channeling is generally expected under these conditions (the incident angle in our experiment is $\pm1^\mr{o}$ off (100)).  Fig. \ref{fig:channeling}(b), on the other hand, shows that the fraction of N ions undergoing channeling is reduced at low energies and is zero for ions below a critical energy.  This critical energy is 0.6 keV for our case.  Thus, channeling will occur for a fraction of the ions implanted at higher energy but not for the lowest energies (namely 0.4 and 0.6 keV) used in our study.

The enhancement of penetration depth for the fraction of ions undergoing channeling can only be estimated for the low kinetic energies used in our study.  Secondary ion mass spectrometry (SIMS) measurements on \N ions implanted into (100) diamond at higher energies (10-30 keV) showed that the deep tail is roughly twice as deep as estimated by SRIM calculations \cite{toyli10}.  Complementary molecular dynamics simulations on 4 keV \N ions predicted a depth of 14 nm (vs. 6.5 nm by SRIM) for channeling ions, underlining the roughly two-fold depth enhancement \cite{nordlund}. 

\subsection{Cluster implantation}
Cluster implantation is a widely used method in semiconductor industry to increase the throughput for shallow implant dopants.  The molecule immediately dissociates when hitting the target surface and the energy of the atoms is given by the mass ratio of the accelerated molecule \cite{liang02}.  For N$_2$ molecules, it is simply one half the implantation energy.  The implantation depth is comparable to monomer ion implantation \cite{liang02}.  However, it is known that cluster implantation reduces the channeling tail but increases the damage \cite{liang02}.  Both effects can affect the yield and thus the photoluminescence signal; this might be related to the slight step in fluorescence we observe between 2.6 keV and 2.2 keV implantation spots [Fig. 1(a) in manuscript].

\section{Photoluminescence measurements}

Photoluminescence and ODMR measurements were carried out on the same home-built inverted confocal microscope, except for curves presented in Fig. 1(c,d), which used a standard fluorescence microscope.  NV defects were excited at 532 nm and emitted photons were filtered at an effective bandwidth of 630-800 nm and collected by an avalanche photodiode. Single center emission was confirmed by photon autocorrelation measurements, ODMR measurements, and the fluorescence intensity.  A 40x, NA=0.95 air objective (Olympus) was used to focus the laser on to diamond sample and collect photons.  The sample was mounted to a motorized three-axis stage (Newport M-462-XYZ-SD) to navigate over the entire 2x2 surface of crystal.  A stationary glass coverslip carrying a thin wire was inserted between objective and diamond sample for microwave excitation.

\begin{figure}[h]
\centering
\includegraphics[width=0.55\textwidth]{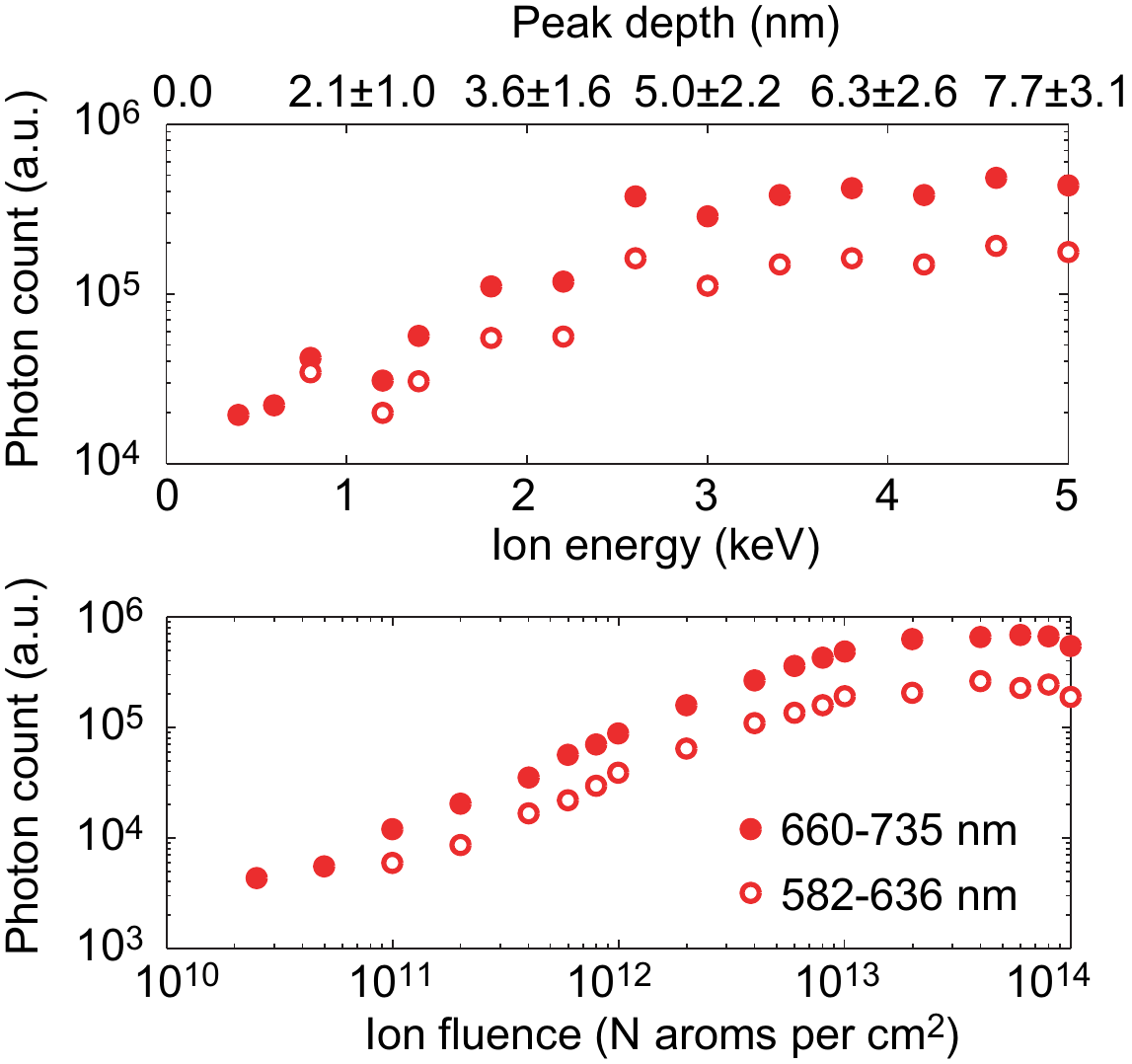}
\caption{\captionstyle Photoluminescence intensity versus energy and fluence, respectively, measured using two sets of filters.}
\label{fig:plfilter}
\end{figure}

\begin{figure}[h]
\centering
\includegraphics[width=0.45\textwidth]{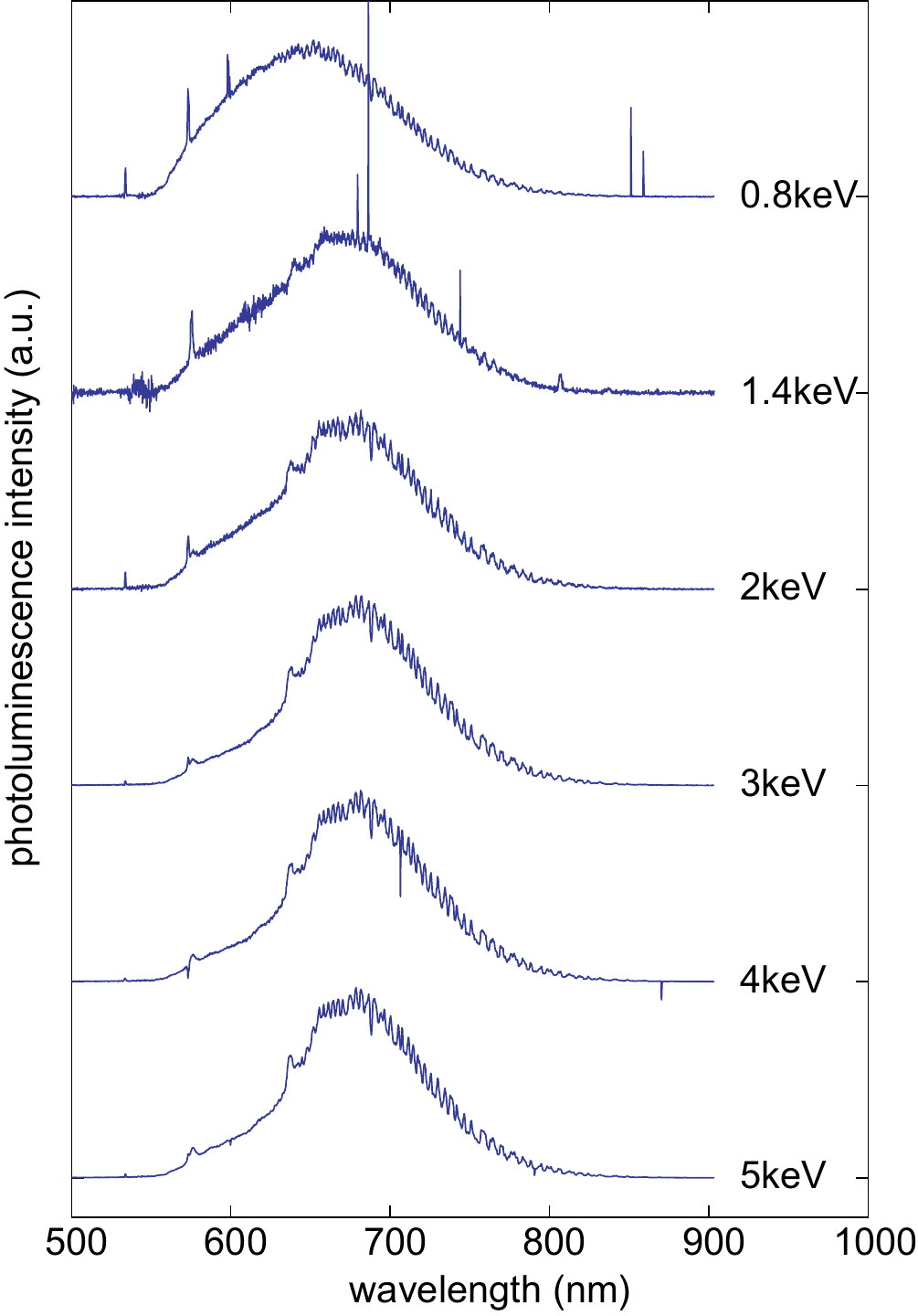}
\caption{\captionstyle Photoluminescence spectra of selected defects at fluence $1\ee{13}\unit{cm^{-2}}$.  Ion energy is given with each curve.  Spectra are vertically offset for clarity.}
\label{fig:plspectra}
\end{figure}

For selective measurements of NV$^0$ and NV$^-$ emission two different sets of filters with bandwidths of 582-636 nm and 660-735 nm, respectively, were used (Fig. \ref{fig:plfilter}). Photoluminescence spectra were also recorded for selected implantation spots to corroborate the findings from the filter measurements (Fig. \ref{fig:plspectra}).  The particular choice of filters is not entirely selective to the two charge states, and the curves shown in Fig. 1(c,d) have been corrected for the overlap of these filters with the excitation spectrum.  The raw curves are shown in Fig. \ref{fig:plfilter}.  Corrected intensities were calculated using the following equation,
\begin{equation}
\left( \begin{array}{c}
I_{\mr{NV^0}} \\
I_{\mr{NV^-}} \\
\end{array} \right)
=
\left( \begin{array}{cc}
0.44 & 0.11 \\
0.26 & 0.66 \\
\end{array} \right)^{-1}
\cdot
\left( \begin{array}{c}
I_{582-636\mr{nm}} \\
I_{660-735\mr{nm}} \\
\end{array} \right),
\end{equation}
where the fraction of photons passing the 582-636 nm filter is 0.44 and 0.11 for NV$^0$ and NV$^-$, respectively, and the fraction of photons passing the 660-735 nm filter is 0.26 and 0.66 for NV$^0$ and NV$^-$, respectively.   Numbers are calculated from the spectra given in Ref. \cite{rondin10}.

Photoluminescence versus energy curves that do not discriminate between NV$^0$ and NV$^-$ were also measured using the confocal setup for the fluences numbers of $8\ee{11}\unit{cm^{-2}}$ and $1\ee{13}\unit{cm^{-2}}$ and gave similar results.

For ODMR linewidth measurements, laser intensity and microwave power was reduced to $\sim1\unit{\mu W}$ and a few $100\unit{kHz}$ Rabi frequency, respectively, to achieve linewidths of $\overset{\sim}{<}1\unit{MHz}$.  This linewidth is still slightly larger than those reported for intrinsically (nuclear \C spin bath) limited high-purity samples and it is likely that some residual power broadening is still present.  The 1 MHz linewidth, however, is sufficient to clearly resolve the hyperfine manifold due to the nitrogen nuclear spin.


\section{Line widths of ODMR spectra}

\subsection{Fitting of spectra}
In order to determine an accurate mean number for the linewidth at various implantation energies, over 30 spectra were collected and individually fitted.  The few spectra that showed a clear, extra hyperfine splitting due to proximal \C nuclei were excluded from statistics.  In the vicinity of each implantation spot, \NNV and \NV spectra were deliberatly taken to ensure that line broadening effects were indeed characteristic to the spot, and not to local variations in the diamond subtrate.  We also measured \NNV spectra at 0.4 keV for different fluences, and found good general agreement (unresolvable lines) in all of them.  This indicates that there is no qualitative difference between ions implanted at different fluence and that there is no dipolar contribution to the linewidth from N donor electron spins.

Line width parameters $\Delta\omega$ ($0.5\times$ the full width at half height) for the ODMR spectra were estimated by fitting a Lorentzian to each of the hyperfine split lines.  For \NNV, the fit function used is
\begin{equation}
I(\omega) = I_0 - I_0 c\left[
\frac{1}{\left(\frac{\omega - \omega_0 + a/2}{\Delta\omega}\right)^2 + 1}
+\frac{1}{\left(\frac{\omega - \omega_0 - a/2}{\Delta\omega}\right)^2 + 1}
\right] .
\end{equation}
Here, $\omega$ is the microwave frequency, and $I_0$, $c$, $x_0$, and $\Delta\omega$ are free fit parameters that correspond to the intensity, the contrast, the center position of the ODMR line, and the linewidth (all in units of angular frequency). $a/(2\pi) =  3.0\unit{MHz}$ is the hyperfine splitting and was assumed a fixed parameter.


\subsection{Calculation of the second moment close to a two-dimensional electronic spin bath}

The second moment of the resonance of an electron spin $S$ due to a quasi-continuous, two-dimensional layer of electron spins $I=1/2$ with density $\rho_A$ (units of spins/m$^2$) is given by
\begin{eqnarray}
\MM & = & \frac{1}{3}\frac{\mu_0^2}{(4\pi)^2}\gamma_I^2\gamma_S^2\hbar^2S(S+1)\int_\mr{layer}dx dy\rho_A\frac{(3\cthsq-1)^2}{r^6} \\
    & = & \frac{1}{4}\frac{\mu_0^2}{(4\pi)^2}\gamma^4\hbar^2\int_\mr{layer}dx dy\rho_A\frac{(3\cthsq-1)^2}{r^6},
\label{eq:m2int}
\end{eqnarray}
where $\gamma_I=\gamma_S=\gamma = 2\pi\cdot 2.8\ee{10}\unit{Hz/T}$ is the electron gyromagnetic ratio, $\mu_0 = 4\pi\ee{-7}\unit{(Vs)/(Am)}$ and $r$ and $\theta$ the distance and angle, respectively, between the NV axis and a surface spin.  This second moment has units of angular frequency squared.
In the following we assume that a small bias field $B_0$ is applied along the NV axis and that all surface spins are aligned with this field.

For integration we note that
\begin{eqnarray}
r(x,y,z) & = & (x^2+y^2+z^2)^{\frac{1}{2}}, \\
\cos(x,y,z) & = & \left[\frac{x}{r} \sin(\theta_0) + \frac{z}{r} \cos(\theta_0)\right]
\end{eqnarray}
where $z=d$ is the distance between spin $S$ and the surface layer, and $\theta_0$ is the angle between NV axis and surface normal. Integration of (\ref{eq:m2int}) gives
\begin{eqnarray}
\MM & = & \frac{1}{4}\frac{\mu_0^2\gamma^4\hbar^2}{(4\pi)^2} \frac{3\pi\rho_A[3+2\cos^2(\theta_0)+3\cos^4(\theta_0)]}{32d^4} , \\
\   & = & \frac{3\mu_0^2\rho_A\gamma^4\hbar^2[3+2\cos^2(\theta_0)+3\cos^4(\theta_0)]}{2048\pi d^4} .
\end{eqnarray}

For our sample, which has a (100) oriented surface, all NV spin will have the same angle set by $\cos\theta_0=1/\sqrt{3}$, or $\theta_0=54.7^\circ$. For this angle, the second moment becomes
\begin{equation}
\MM_{\theta_0=54.7^\circ} = \frac{3\mu_0^2\rho_A\gamma^4\hbar^2}{512\pi d^4}.
\end{equation}
%

\subsection{Exclusion of line broadening by \NN spin bath}
We have also calculated the average distance between \NN defects in the implanted diamond in order to exclude extra line broadening due to the N donor electron spin bath.  ODMR spectra were recorded at a single fluence of $8\ee{11}\unit{cm^2}$ and on defects at the perimeter of the implantation spots, where the density is reduced by about $<1/100$ as estimated by fluorescence intensity.  Peripheral defects were chosen for all measurements in order to resolve individual luminescent centers and to ensure N ions were far enough apart.  From the ion density we calculate the average (median) N-N distance to $d\approx0.5\sqrt{0.01 8\ee{11}\unit{cm^2}}\approx 50\unit{nm}$.  The dipolar interaction at this distance is about
\begin{equation}
\omega_\mr{D} \approx \frac{\mu_0}{4\pi} \frac{1}{d^3}\frac{\hbar\gamma^2}{4} \approx 2\pi \cdot 100\unit{Hz},
\end{equation}
where $\gamma$ is the electron gyromagnetic ratio.  This value is much less than the base linewidth of $\sim1\unit{MHz}$ relevant to our study.

\section{Spin echo measurements and fitting of decay times}

Coherence times $T_2$ were measured using a Hahn echo sequence with equal free evolution times $\tau' = \tau$ before and after the central $\pi$ pulse.  This pulse sequence produces the decay of the echo maximum.  The echo decay curves were then fitted by the following equation (adapted from Ref. \cite{childress06}),
\begin{equation}
I(\tau) = A \exp\left[-\frac{(2\tau)^{n_1}}{T_2^3}\right]  \cdot \sum_{k=0}^N { \exp\left[-\frac{(\tau-k\tau_{r})^{n_2}}{(2 \tau_c)^2}\right] } + C,
\end{equation}
where $\tau_c$ describes the fast initial decay caused by the fluctuating \C nuclear spin bath, $T_2$ the slower decay of the ``echo revivals'' appearing at the periodicity $\tau_r$ of the \C Larmor precession, and $n_1=3$, $n_2=2$.  Here, $A$, $C$, $\tau_c$, $T_2$ and $\tau_r$ are free fit parameters, and $N$ was adjusted to match the number of revivals seen.  We also tried fitting with arbitrary exponentials $n_1$, $n_2$, but fits would either not converge or yielded the same values for $\tau_c$ and $T_2$ within experimental error.  Thus, no conclusion can be drawn on the exponent.  In principle, the $T_2$ decay exponential will change from $n_1=3$ to $n_1=1$ for a rapidly fluctuating environment (such as caused by fast reorientation of surface spins), but from the present data we cannot favor one over the other.
